\documentclass[11pt,showpacs,preprintnumbers,amsmath,aps,amssymb]{revtex4}
\usepackage{graphicx}
\usepackage{dcolumn}
\usepackage{bm}
\begin{document}

\title{Cosmological Bianchi Class A models in S\'aez-Ballester theory}

\author{J. Socorro$^1$}
\email{socorro@fisica.ugto.mx}
\author{Paulo A. Rodr\'iguez$^1$}
\email{paulo999@yahoo.com}
\author{Abraham Espinoza-Garc\'ia$^1$}
\email{abraham@fisica.ugto.mx}
\author{Luis O. Pimentel}
\email{lopr@xanum.uam.mx}
\author{Priscila Romero}
\email{prisscience@gmail.com}
 
\affiliation{$^1$Departamento de F\'{\i}sica de la  DCeI de la  Universidad de 
Guanajuato-Campus Le\'on\\
A.P. E-143, C.P. 37150,  Guanajuato, M\'exico\\
$^2$Departamento de F\'{\i}sica de la  Universidad Aut\'onoma Metropolitana\\
Apartado Postal 55-534, 09340, M\'exico, D.F.\\
$^2$Facultad de Ciencias de la Universidad Aut\'onoma del Estado de 
M\'exico,\\
Instituto Literario  No. 100, Toluca, C.P. 50000,  Edo de Mex}


\begin{abstract}
We use the  S\'aez-Ballester (SB) theory on  anisotropic Bianchi Class A cosmological model, 
with barotropic fluid and cosmological constant, using the Hamilton or Hamilton-Jacobi approach. Contrary to claims in the
  specialized literature, it is shown that the Sáez-Ballester theory cannot
provide a realistic solution to the dark matter problem of Cosmology for the dust epoch, without a fine tunning because 
the contribution of the scalar field in this theory is equivalent to a stiff fluid (as can be seen from the 
energy--momentum tensor for the scalar field), that evolves in a different way as the dust component. 
To have similar contributions of the scalar component and the dust component implies that their past values were fine tunned. 
So, we reinterpreting this null result as an indication that dark matter plays a central role in the formation of structures and galaxy evolution,  
having measureable effects in the cosmic microwave bound radiation, and than this formalism yield to this epoch as primigenius results.
 We do the mention that this formalism was used recently in the so called K-essence theory 
 applied to dark energy problem, in place to the dark matter problem. Also,
we include a quantization procedure of the theory which can be simplified by 
  reinterpreting the theory in the Einstein frame, where the scalar field can be interpreted
   as part of the matter content of the theory, and exact solutions to the Wheeler-DeWitt equation are found, employing
   the Bianchi Class A cosmological models. 
  \end{abstract}

\pacs{02.30.Jr; 04.60.Kz; 12.60.Jv; 98.80.Qc.}
 \maketitle                            

\section{Introduction}
Several observations suggest that in galaxies and galaxy clusters there is an important quantity of matter that is not interacting  
electromagnetically, but only through gravitation. This is the well known dark matter problem. Several solutions have been consider 
for this problem, modifying the gravitational theory or introducing new forms of matter and interaccions. To adress the dark matter
 problem Saez and Ballester (SB) \citep{s-b} formulated a scalar-tensor theory of gravitation
in which the metric is coupled with a dimensionless scalar field. In  a recent analysis using the standard scalar field cosmological models
\citep{socorro1,socorro2}, contrary to claims in the specialized literature,
it is shown that the SB theory cannot provide a realistic solution to the dark matter problem
of Cosmology for the dust epoch, because the contribution of the scalar field is equivalent to stiff matter.
 We can reinterpret this result in a sense
  that the  galaxy halo was formed during this primigenius epoch and its evolution until the dust era using the
  standard scalar field cosmological theory. In  this theory the strength of the coupling between gravity and the scalar field
is determined by an arbitrary coupling constant $\omega$. 
This constant $\omega$ can be used to have a lorenzian (-1,1,1,1) or seudo-lorenzian (-1,-1,1,1) signature when we build the
Wheeler-DeWitt equation. The values for this constant,  in the classical regime, are dictated by the condition to have real functions.
 Other problem inherent to this theory  is that
not exist how build the invariants with this field as in the case to scalar curvature. So, was necessary to reinterpret the formalism where
this field is considered as matter content in the theory in the Einstein frame.

   On the other hand, this approach is classified 
 with another name, by instant, Armendariz-Picon et al, called this formalism as  K-essence \citep{armendariz},
 as a dynamical solution for explaining naturally why the universe has entered an epoch of accelerated expansion at a late stage
of its evolution. Instead, K-essence is based on the idea of a dynamical attractor solution which causes it to
act as a cosmological constant only at the onset of matter domination.
Consequently, K-essence overtakes the matter density and induces cosmic acceleration at about the present
epoch.  Usually K-essence models
are restricted to the Lagrangian density of the form
\begin{equation}
\rm S=\int d^4x \, \sqrt{-g}\, f(\phi) \, \left(\nabla \phi \right)^2.
\end{equation}
One of the motivations to consider this type of Lagrangian originates from string theory \-  \citep{string}. 
For more details for K-essence applied
to dark energy,  you can  see in \citep{copeland} and reference therein.  Many works in SB formalism 
in the classical regime have been done, where
the Einstein field equation is solved in a direct way, using a particular ansatz for the main scalar factor of the universe
\citep{singh-a,shri-a,mohanty-a,singh-shri}, 
yet a study of the aniso\-tro\-py behaviour trough the form  introduced in 
the line element has been conected \citep{reddy1,mohanty-b,mohanty-c,adhav,rao-a,rao-b,shri-b,tripa,singh-b,pradan}. 
  
On another front, the quantization program of this theory has not been constructed. 
The main complication can be traced to the lack of an ADM type formalism. We can transform this theory to conventional one
 where the dimensionless 
scalar field is obtained from
energy-momentum tensor as an exotic matter contribution, and in this sense we can use this formalism for 
the quantization program, where the ADM formalism is well
known \citep{ryan1}.

In this work, we use this formulation to obtain classical and quantum exact solutions to
anisotropic Bianchi Class A cosmological models with  stiff matter. 
 The first step is to write SB formalism in the usual manner, that is, we calculate the
corresponding energy-momentum tensor to the scalar field and give the equivalent Lagrangian density. Next, we proceed to obtain
the corresponding canonical Lagrangian ${\cal L}_{can}$ to  Bianchi Class A cosmological models through the Legendre transformation, 
we calculate the classical Hamiltonian 
${\cal H}$, from which we find the Wheeler-DeWitt (WDW) equation of the corresponding cosmological model under study. We employ in this work
the Misner parametrization due that a natural way appear the anisotropy parameters to the scale factors.

The simpler generalization to Lagrangian density for the SB theory  \citep{s-b} with the cosmological term, is  
\begin{equation}
\rm {\cal L}_{geo}=\left( R- 2\Lambda - F(\phi) \phi_{,\gamma} \phi^{,\gamma}\right), \label{lagrangian}
\end{equation}
where $\phi^{,\gamma}=g^{\gamma \alpha} \phi_{,\alpha}$, R the scalar curvature, $F(\phi)$ a dimensionless function of the  scalar field.. 
In classical field theory with scalar field, this formalism corresponds to null potencial in the field $\phi$, but the
kinetic term is exotic by the factor $F(\phi)$. 

From the Lagrangian (\ref{lagrangian}) we can build the complete action
\begin{equation}
\rm I=\int_{\Sigma} \sqrt{-g}({\cal L}_{geo}+ {\cal L}_{mat})d^4x,\label{action}
\end{equation}
where ${\cal L}_{mat}$ is the matter Lagrangian,  g is the determinant of metric tensor.
The field equations for this theory are

\begin{subequations}
\begin{eqnarray}
\rm G_{\alpha \beta}+g_{\alpha\beta}\Lambda-F(\phi) \left( \phi_{,\alpha}\phi_{,\beta}
 - \frac{1}{2} g_{\alpha \beta} \phi_{,\gamma} \phi^{,\gamma} \right)
&=& \rm - 8\pi G T_{\alpha \beta}, \label{efe} \\
\rm 2F(\phi) \phi^{,\alpha}_{\,\,;\alpha} + \frac{dF}{d\phi}\phi_{,\gamma} \phi^{,\gamma}&=&0, \label{fe}
\end{eqnarray}
\end{subequations}
 where G is the gravitational constant and as usual the semicolon means a covariant derivative.

 The equation (\ref{fe}) take the following form for all cosmological Bianchi Class A models, assuming that the scalar field is 
 only time dependent ( here $\prime=\frac{d}{d\tau}=\frac{d}{Ndt}$)
 $$\rm 3\Omega^\prime \phi^\prime F+ \phi^{\prime\prime}F +\frac{1}{2}\frac{dF}{d\phi} \phi^{\prime 2}=0$$
which can be put in quadrature form as
\begin{equation}
\rm \frac{1}{2}F\phi^{\prime 2}= F_0 e^{-6\Omega}, \label{phifield}
\end{equation}
this equation is seen  as corresponding to a stiff matter content contribution. 

The same set of equations(\ref{efe},\ref{fe}) is obtained if we consider the
scalar field $\phi$ as part of the matter budget, i.e. say 
$\rm {\cal L}_{\phi}=\rm - F(\phi) g^{\alpha \beta}\phi_{,\alpha} \phi_{,\beta}$
with your 
corresponding energy-momentum tensor 
\begin{equation}
\rm T_{\alpha \beta}=F(\phi) \left( \phi_{,\alpha}\phi_{,\beta}
 - \frac{1}{2} g_{\alpha \beta} \phi_{,\gamma} \phi^{,\gamma} \right). \label{ener-mom}
 \end{equation}
 which is conserved directly considering a stiff matter era in a barotropic scalar fluid (see appendix section 8).
In this new line of reasoning, 
 action (\ref{action}) can be rewritten as a geometrical part (Hilbert-Einstein with $\Lambda$) 
 and matter content (usual matter plus a term that corresponds to the exotic scalar field component of SB theory).

In this way, we write the action (\ref{action}) in the usual form
\begin{equation}
\rm I=\int_{\Sigma} \sqrt{-g}\left( R- 2\Lambda +{\cal L}_{mat}+{\cal L}_\phi \right)d^4x,\label{action1}
\end{equation}
and consequently, the classical equivalence between the two theories. We can infer that this correspondence also is satisfied in the quantum 
regime, so we can use this structure for the quantization program, where the ADM formalism is well
known for different classes of matter \citep{ryan1}. Using this action we obtain the Hamiltonian for SB. 
We find that the WDW equation is solved when we choose one ansatz similar to this employed in the Bohmian formalism 
 of quantum mechanics and the gravitational part in the solutions are the same that these found in the literature, years ago \citep{os}. 

This work is arranged as follow. In section 3 we present the method used, employing the FRW cosmological model with barotropic perfect fluid
and cosmological constant. In section 4 we construct the Lagrangian and Hamiltonian densities for 
the anisotropic Bianchi Class A cosmological model. In
section 5  the classical  solutions using the Jacobi formalism are found. Here we present partial results 
in the solutions for some Bianchi's cosmological models. Classical solution to Bianchi I is complete in any gauge, but the
Bianchi II and $\rm VI_{h=-1}$, the solutions are found in particular gauge. Other Biachi's, only the master equation are presented. 
In Section 6 the complete cuantization scheme is presented, obtaining the corresponding Wheeler-DeWitt equation and its solutions 
are presented in unified way using the classification scheme of Ellis and MacCallum \citep{ellis} and
Ryan and Shepley, \citep{ryan-she}.

\section{The method}
Let us start with the line element for a homogeneous and isotropic FRW
universe  
\begin{equation}
  ds^2= -N^2(t) dt^2 + a^2(t) \left[ \frac{dr^2}{1-\kappa r^2} + r^2
    d\Omega^2 \right] \, , \label{frw}
\end{equation}
where $a(t)$ is the scale factor, $N(t)$ is the lapse function, and
$\kappa$ is the curvature constant that can  take the values $0$,
$1$ and $-1$, for flat, closed and open universe, respectively. The
total Lagrangian density then reads
\begin{equation}
  {\cal L} = \frac{6\dot a^2 a}{N} - 6\kappa N a + \frac{F(\phi)
    a^3}{N} \dot \phi^2 + 16\pi G N a^3 \rho - 2N a^3 \Lambda \,
  , \label{frw-lagrangian}
\end{equation}
where $\rho$ is the energy density of matter, we will assume that it     
complies with a barotropic equation of state of the form $p=\gamma
\rho$, where $\gamma$ is a constant.
The matter content is assumed as a perfect fluid $\rm T_{\mu\nu}= (\rho + p)u_\mu u_\nu + g_{\mu\nu} p$
where $u_\mu$ is the fluid four-velocity  satisfying $\rm u_\mu u^\mu=-1$ . Taking the covariant derivative we obtain
the relation
$$
3\dot \Omega \rho +3\dot \Omega p+\dot \rho=0,
$$
whose solution becomes
\begin{equation}
\rm \rho =\rho_\gamma e^{-3\Omega \left( 1+\gamma \right)}.
 \label{density}
\end{equation}
where $\rho_\gamma$ is an integration constant.
 
From the  canonical form of  the Lagrangian density (\ref{frw-lagrangian}), and the 
 solution for the barotropic fluid equation of motion, we find  the Hamiltonian
density for this theory, where the momenta are defined in the usual way 
$\Pi _{q^i}=\rm \frac{\partial {\cal L} }{\partial \dot q^i}$, where $\rm q^i=(a, \phi)$ are
the field coordinates for this system,
\begin{eqnarray}
  \Pi_a &=& \frac{\partial {\cal L}}{\partial \dot a} = \frac{12a\dot
    a}{N}, \qquad \rightarrow \qquad \dot a = \frac{N\Pi_a}{12 a} \, ,
  \nonumber \\
  \Pi_\phi &=& \frac{\partial {\cal L}}{\partial \dot \phi} = \frac{2F
    a^3 \dot{\phi}}{N} \, , \qquad \rightarrow \qquad \dot{\phi} =
  \frac{N\Pi_\phi}{2F a^3 } \, , \label{momentasq}
\end{eqnarray}
so, the Hamiltonian density become
\begin{equation}
  {\cal H} = \frac{a^{-3}}{24} \left[ a^2 \Pi_a^2 + \frac{6}{F(\phi)}
    \Pi_\phi^2 + 144 \kappa a^4 + 48a^6 \Lambda -384 \pi G \rho_\gamma
    a^{3(1-\gamma)} \right] . \label{hamiltonian}
\end{equation}

Using the transformation $\Pi_q=\frac{d S_q}{d q}$, the
Einstein-Hamilton-Jacobi (EHJ) associated to Eq.~(\ref{hamiltonian}) is
\begin{equation}
  a^2 \left( \frac{d S_a}{d a} \right)^2 + \frac{6}{F(\phi)}
  \left(\frac{d S_\phi}{d \phi} \right)^2 + 48a^6 \Lambda - 384 \pi G
  \rho_\gamma a^{3(1-\gamma)} = 0 \, ,.
\end{equation}
The EHJ equation can be further separated in the equations
\begin{eqnarray}
  \frac{6}{F(\phi)} \left( \frac{d S_\phi}{d \phi} \right)^2 &=& \mu^2
  \, , \label{pphi} \\
  a^2 \left( \frac{d S_a}{d a} \right)^2  + 48a^6 \Lambda - 384 \pi G
  \rho_\gamma a^{3(1-\gamma)} &=& -\mu^2 \, ,  \label{aa}
\end{eqnarray}
where $\mu$ is a separation constant. With the help of
Eqs.~(\ref{momentasq}), we can obtain the solution up to quadratures of
Eqs.~(\ref{pphi}) and~(\ref{aa}),
\begin{subequations}
\begin{eqnarray}
  \int \sqrt{F(\phi)} \, d\phi &=& \frac{\mu}{2\sqrt{6}}\int a^{-3}
  (\tau) \, d\tau \, , \label{phi-new} \\
  \Delta \tau &=& \int \frac{a^2da}{\sqrt{\frac{8}{3}\pi G \rho_\gamma
      a^{3(1-\gamma)} - \frac{\Lambda}{3}a^6 - \nu^2}} \,
  , \label{aa-new}
\end{eqnarray}
\end{subequations}
with $\nu=\frac{\mu}{12}$.
Eq.~(\ref{phi-new}) readily indicates that
\begin{equation}
  F(\phi) \dot \phi^2 = 6\nu^2 a^{-6}(\tau) \, . \label{missing}
\end{equation}
 Also, this equation could be obtained by 
solving equation (\ref{fe}). Moreover,
the matter contribution of the SB scalar field to the r.h.s. of the
Einstein equations would be
\begin{equation}
  \rho_\phi = \frac{1}{2} F(\phi) \dot{\phi}^2 \propto a^{-6} \, .
\end{equation}
this energy density of a scalar field has the range of scaling behaviors \citep{andrew,ferreira}, is say, scales
exactly as a power of the scale factor like,  $\rho_\phi\propto a^{-m}$, when the dominant component has an energy
density which scales as similar way.
So, the contribution of the scalar field is the same as that of
stiff matter with a barotropic equation of state $\gamma = 1$.
This is an interesting result, since the original SB theory was thought of as
a way to solve the missing matter problem 
now generically called the dark matter problem.  To solve the latter,
one needs a fluid behaving as dust with $\gamma = 0$, it is surprising that
such a general result remains unnoticed until now in the literature
about SB.  This is an instance of the results of the analysis of the energy momentum tensor of a scalar
field by Marden \citep{marden} for General Relativity with scalar matter and by Pimentel \citep{pimentel}
for the general scalar tensor theory. In both works a free scalar field is equivalent to a stiff matter fluid.

Furthermore, having  identified the general evolution of the scalar field with that of a stiff
fluid means that the Eq. (\ref{aa-new}) can be integrated separately without a complete solution for
the scalar field. In \citep{socorro2} appear a compilation of exact
solutions in the case of the original SB theory to FRW cosmological model  and in \citep{socorro1} were presented
the classical and quantum solution to Bianchi type I.

\section{The master Hamiltonian to Bianchi Class A cosmological models}
Let us recall here the canonical formulation in the ADM formalism of the 
diagonal Bianchi Class A cosmological models. The metric has the form
\begin{equation}
\rm ds^2= -dt^2 + e^{2\Omega(t)}\, (e^{2\beta(t)})_{ij}\, \omega^i \, 
\omega^j, \label {met}
\end{equation}
where $\rm \beta_{ij}(t)$ is a 3x3 diagonal
matrix, $\rm \beta_{ij}= diag(\beta_++ \sqrt{3} \beta_-,\beta_+- \sqrt{3} \beta_-, -2\beta_+)$,
$\Omega(t)$ is a scalar and $\rm \omega^i$ are one-forms that  characterize  each cosmological Bianchi
type model, and that obey 
$\rm d\omega^i= \frac{1}{2} C^i_{jk} \omega^j \wedge \omega^k,$
$\rm C^i_{jk}$ the structure constants of the corresponding invariance
group, these are included in table 1.

\vglue 1cm
\begin{center}
\begin{tabular}{|c|l|} \hline
 {\rm Bianchi type}&1-forms $\omega^i$\\ \hline
I  & $\rm \omega^1=dx^1$,\quad $\rm \omega^2=dx^2$, \quad $\rm \omega^3=dx^3$ \\ \hline
II & $\rm \omega^1=dx^2 - x^1 dx^3$,\quad $\rm \omega^2=dx^3$,\qquad $\rm \omega^3=dx^1$ \\ \hline
$\rm VI_{h=-1}$ & $\rm \omega^1=e^{-x^1}dx^2$,\qquad $\rm \omega^2=e^{x^1}dx^3$,\qquad $\rm \omega^3=dx^1$\\ \hline
$\rm VII_0$ & $\rm \omega^1=dx^2 + dx^3$,\qquad $\rm \omega^2=-dx^2 + dx^3$, \qquad $\rm \omega^3=dx^1$\\ \hline
VIII & $\rm \omega^1=dx^1 + [1 + (x^1)^2]dx^2 +[x^1 - x^2 - (x^1)^2 x^2]dx^3$,\\
 & $\rm \omega^2=2x^1 dx^2+ (1-2x^1 x^2)dx^3$, \\ 
   & $\omega^3=dx^1 + [-1 + (x^1)^2]dx^2 +[x^1 + x^2 - (x^1)^2 x^2]dx^3$ \\ \hline
IX  & $\rm \omega^1= -\sin(x^3)dx^1  + \sin(x^1)\cos(x^3)dx^2$,\\ 
&$\rm \omega^2=\cos(x^3)dx^1+ \sin(x^1)\sin(x^3)dx^2$, \quad
  $\rm \omega^3=\cos(x^1)dx^2 + dx^3$ \\ \hline
\end{tabular}
\vglue .5cm
Table 1. \emph{one-forms for the Bianchi Class A models.}
\end{center}

We use the Bianchi type IX cosmological model as toy model to apply method discussed in the  previous section.
 The total Lagrangian density then reads
\begin{eqnarray}
 \rm {\cal L}_{_{IX}} &=&\rm  e^{3\Omega} \left[6 \frac{\dot \Omega^2 }{N} - 6 \frac{\dot \beta_+^2}{N}- 6 \frac{\dot \beta_-^2}{N} 
  + \frac{F(\phi)}{N} \dot \phi^2 + 16\pi G N  \rho - 2N  \Lambda \right.  \nonumber\\
 &&\rm \left. +Ne^{-2\Omega}\left\{\frac{1}{2}\left(e^{4\beta_++4\sqrt{3}\beta_-} +e^{4\beta_+-4\sqrt{3}\beta_-}+ e^{-8\beta_+} \right) 
 \right. \right.  \nonumber\\ 
  &&\rm \left. \left. 
  -\left(e^{-2\beta_++2\sqrt{3}\beta_-}+e^{-2\beta_+-2\sqrt{3}\beta_-}+e^{4\beta_+} \right) \right\}\right] , \label{ix-lagrangian}
\end{eqnarray}
making the calculation of momenta in the usual way, $\rm \Pi_{q^\mu}=\frac{\partial{\cal L}}{\partial{\dot q^\mu}}$, where 
$\rm q^{\mu}=(\Omega, \beta_+,\beta_-, \phi)$
\begin{eqnarray*}
\rm \Pi_\Omega&=&\rm \frac{12}{N}e^{3\Omega}\dot \Omega, \quad \rightarrow \quad \dot \Omega=\frac{N}{12}e^{-3\Omega}\Pi_\Omega \\
\rm \Pi_+&=& \rm -\frac{12}{N}e^{3\Omega}\dot \beta_+, \quad \rightarrow \quad \dot \beta_+=-\frac{N}{12}e^{-3\Omega}\Pi_+\\
\rm \Pi_-&=& \rm -\frac{12}{N}e^{3\Omega}\dot \beta_-, \quad \rightarrow \quad \dot \beta_-=-\frac{N}{12}e^{-3\Omega}\Pi_+\\
\rm \Pi_\phi&=& \rm \frac{2F}{N}e^{3\Omega}\dot \phi, \quad \rightarrow \quad \dot \phi=\frac{N}{2F}e^{-3\Omega}\Pi_\phi\\
\end{eqnarray*}
and introducing into the Lagrangian density, we obtain the canonical Lagrangian as 
$$\rm {\cal L}_{_{IX}}= \Pi_{q^\mu} \dot q^\mu -N {\cal H}_\perp,$$
with the Hamiltonian density
\begin{equation}
{\cal H}_\perp= \rm \frac{e^{-3\Omega}}{24}\left(-\Pi^2_\Omega -\frac{6}{F(\phi)}\Pi_\phi^2+ \Pi^2_+ +\Pi^2_- +U(\Omega,\beta_\pm)+C_1\right ), 
\label {ham}
\end{equation}
where the gravitational potential becomes,
$$\rm U(\Omega,\beta_\pm)= 12e^{4\Omega}\left(e^{4\beta_++4\sqrt{3}\beta_-}+e^{4\beta_+-4\sqrt{3}\beta_-}+ e^{4\beta_+}
-2\left\{e^{4\beta_+}+e^{2\beta_+-2\sqrt{3}\beta_-}+e^{-2\beta_++2\sqrt{3}\beta_-}\right\}\right),$$
with $\rm C_1=384\pi G \rho_1$ corresponding to stiff matter epoch, $\gamma=1$.

The equation (\ref{ham}) can be considered as a master equation for all Bianchi Class A cosmological 
model in the stiff epoch in the S\'aez-Ballester theory, with 
$\rm U(\Omega,\beta_\pm)$  is the potential term of the 
cosmological model under consideration, that can read it to table II. 
\vglue 1cm
\begin{table}[h]
\begin{center}
\begin{tabular}{|c|l|} \hline
 {\rm Bianchi type}& Hamiltonian density ${\cal H}$\\ \hline
I  & $\rm \frac{e^{-3\Omega}}{24}\left[-\Pi_\Omega^2-\frac{6}{F}\Pi^2_\phi +\Pi_+^2+\Pi_-^2-48\Lambda e^{6\Omega}+384\pi G \rho_\gamma 
e^{-3(\gamma-1)\Omega} \right]$ \\ \hline
II & $\rm \frac{e^{-3\Omega}}{24}\left[-\Pi_\Omega^2-\frac{6}{F}\Pi^2_\phi +\Pi_+^2+\Pi_-^2-48\Lambda e^{6\Omega}+384\pi G \rho_\gamma 
e^{-3(\gamma-1)\Omega} \right.$ \\
& \qquad $\rm \left. +12e^{4\Omega} e^{4\beta_++4\sqrt{3}\beta_-}\right]$ \\ \hline
${\rm VI_{-1}}$ & $\rm \frac{e^{-3\Omega}}{24}\left[-\Pi_\Omega^2-\frac{6}{F}\Pi^2_\phi +\Pi_+^2+\Pi_-^2-48\Lambda e^{6\Omega}+384\pi G \rho_\gamma 
e^{-3(\gamma-1)\Omega} \right.$ \\
& \qquad $\rm \left. + 48 e^{4\Omega}e^{4\beta_+}\right]$\\ \hline
${\rm VII_{0}}$ & $\frac{e^{-3\Omega}}{24}\left[-\Pi_\Omega^2-\frac{6}{F}\Pi^2_\phi +\Pi_+^2+\Pi_-^2-48\Lambda e^{6\Omega}+384\pi G \rho_\gamma 
e^{-3(\gamma-1)\Omega} \right. $\\
&$ \qquad \left.+12e^{4\Omega}\left(e^{4\beta_++4\sqrt{3}\beta_-}- e^{4\beta_+}+e^{4\beta_+-4\sqrt{3}\beta_-}\right)\right]$\\ \hline
VIII & $\rm \frac{e^{-3\Omega}}{24}\left[-\Pi_\Omega^2-\frac{6}{F}\Pi^2_\phi +\Pi_+^2+\Pi_-^2-48\Lambda e^{6\Omega}+384\pi G \rho_\gamma 
e^{-3(\gamma-1)\Omega} \right. $\\
&$\qquad  \left.+12e^{4\Omega}\left(e^{4\beta_++4\sqrt{3}\beta_-}+e^{4\beta_+-4\sqrt{3}\beta_-}+ e^{-8\beta_+} \right. \right. $\\
&$\rm \qquad \left.\left. -2\left\{e^{4\beta_+}-e^{-2\beta_+-2\sqrt{3}\beta_-}-e^{-2\beta_++2\sqrt{3}\beta_-}\right\}\right)\right]$\\ \hline
IX  & $\rm \frac{e^{-3\Omega}}{24}\left[-\Pi_\Omega^2-\frac{6}{F}\Pi^2_\phi +\Pi_+^2+\Pi_-^2-48\Lambda e^{6\Omega}+384\pi G \rho_\gamma 
e^{-3(\gamma-1)\Omega} \right. $\\
&$\qquad  \left.+ 12e^{4\Omega}\left(e^{4\beta_++4\sqrt{3}\beta_-}+e^{4\beta_+-4\sqrt{3}\beta_-}+ e^{-8\beta_+} \right.\right.$ \\
& $\rm \qquad  \left.\left. -2\left\{e^{4\beta_+}+e^{2\beta_+-2\sqrt{3}\beta_-}+e^{-2\beta_++2\sqrt{3}\beta_-}\right\}\right)\right]$ \\ \hline
\end{tabular}
\vglue .5cm
Table II. \emph{ Hamiltonian density for the Bianchi Class A models.}
\end{center}
\end{table}
\section{classical scheme}
In this section, we present the classical solutions to all Bianchi Class A cosmological models using the appropriate set of variables,
\begin{eqnarray}
\rm \beta_1&=&\rm \Omega+\beta_+ +\sqrt{3} \beta_-,\nonumber\\
\rm \beta_2&=&\rm \Omega+\beta_+ -\sqrt{3} \beta_-,\nonumber\\
\rm \beta_3&=&\rm \Omega-2\beta_+. \label{transformation}
\end{eqnarray} 

\subsection{Bianchi I}
For building one master equation for all Bianchi Class A models, we begin with the simplest model give by the Bianchi I, 
and give the general treatment. 
 The corresponding Lagrangian for this cosmological model is written as
\begin{equation}
\rm {\cal L}_I=e^{\beta_1+\beta_2+\beta_3}\left[\frac{2\dot \beta_1 \dot \beta_2}{N}+\frac{2\dot \beta_1 \dot \beta_3}{N}
+\frac{2\dot \beta_2 \dot \beta_3}{N} +\frac{F(\phi)\dot\phi^2}{N} +16N\pi G \rho_\gamma \,e^{-(1+\gamma)(\beta_1+\beta_2+\beta_3)}-
2N\Lambda \right],
\end{equation} 
the momenta associated to the variables $(\beta_i,\phi)$ are
\begin{eqnarray}
\rm \Pi_1&=&\rm \frac{2}{N}(\dot \beta_2 + \dot \beta_3)e^{\beta_1+\beta_2+\beta_3}, 
\qquad \dot \beta_1= \frac{N}{4}\,e^{-(\beta_1+\beta_2+\beta_3)} (\Pi_2+\Pi_3-\Pi_1),\nonumber\\
\rm \Pi_2&=&\rm \frac{2}{N}(\dot \beta_1 + \dot \beta_3)e^{\beta_1+\beta_2+\beta_3},
\qquad \dot \beta_2= \frac{N}{4}\,e^{-(\beta_1+\beta_2+\beta_3)} (\Pi_1+\Pi_3-\Pi_2),\nonumber\\
\rm \Pi_3&=&\rm \frac{2}{N}(\dot \beta_1 + \dot \beta_2)e^{\beta_1+\beta_2+\beta_3},
\qquad \dot \beta_3= \frac{N}{4}\,e^{-(\beta_1+\beta_2+\beta_3)} (\Pi_1+\Pi_2-\Pi_3),\nonumber\\
\rm \Pi_\phi&=&\rm \frac{2F\dot \phi}{N}e^{\beta_1+\beta_2+\beta_3}, \qquad \dot \phi=\frac{N}{2F}e^{-(\beta_1+\beta_2+\beta_3)}\Pi_\phi.
\label{uno}
\end{eqnarray}
so, the Hamiltonian is
\begin{eqnarray}
\rm {\cal H}_I&=&\rm \frac{1}{8}e^{-(\beta_1+\beta_2+\beta_3)} \left[-\Pi_1^2-\Pi_2^2-\Pi_3^2 +\frac{2}{F}\Pi_\phi^2+2\Pi_1 \Pi_2
+2\Pi_1 \Pi_3 +2\Pi_2 \Pi_3 \right. \nonumber\\
&&\rm \left. +16\Lambda e^{2(\beta_1+\beta_2+\beta_3)}-128\pi G \rho_\gamma e^{(1-\gamma)(\beta_1+\beta_2+\beta_3)}\right], \label{hami}
\end{eqnarray}
using the hamilton equation, where $\prime=\frac{d}{d\tau}=\frac{d}{Ndt}$, we have 
\begin{eqnarray}
\rm  \Pi_1^\prime&=& \rm -4\Lambda e^{\beta_1+\beta_2+\beta_3}+16\pi G(1-\gamma)\rho_\gamma e^{-\gamma(\beta_1+\beta_2 +\beta_3)}, \label{p1i} \\
\rm  \Pi_2^\prime&=& \rm -4\Lambda e^{\beta_1+\beta_2+\beta_3}+16\pi G(1-\gamma)\rho_\gamma e^{-\gamma(\beta_1+\beta_2 +\beta_3)}, \label{p2i}\\
\rm \Pi_3^\prime&=& \rm -4\Lambda e^{\beta_1+\beta_2+\beta_3}+16\pi G(1-\gamma)\rho_\gamma e^{-\gamma(\beta_1+\beta_2 +\beta_3)}, \label{p3i}\\
\rm  \Pi_\phi^\prime &=& \rm \frac{1}{4}e^{-(\beta_1+\beta_2+\beta_3)}\, \frac{ F^\prime}{F^2  \phi^\prime}\Pi_\phi^2 , \label{p4i}\\
\rm  \beta_1^\prime&=&\rm \frac{1}{4}e^{-(\beta_1+\beta_2+\beta_3)}\left[-\Pi_1 +\Pi_2+\Pi_3 \right], \label{beta1}\\
\rm  \beta_2^\prime&=&\rm \frac{1}{4}e^{-(\beta_1+\beta_2+\beta_3)}\left[-\Pi_2 +\Pi_1+\Pi_3 \right], \label{beta2}\\
\rm  \beta_3^\prime&=&\rm \frac{1}{4}e^{-(\beta_1+\beta_2+\beta_3)}\left[-\Pi_3 +\Pi_1+\Pi_2 \right], \label{beta3}\\
\rm  \phi^\prime&=& \rm \frac{1}{2F}e^{-(\beta_1+\beta_2+\beta_3)}\Pi_\phi. \label{phi1i}
\end{eqnarray}
equations (\ref{p1i},\ref{p2i},\ref{p3i}) implies 
\begin{equation}
\rm  \Pi_1=\Pi_2+k_1=\Pi_3+k_2. \label{momentas}
\end{equation}
Also, the differential equation for field $\phi$ can be reduced to quadratures when we use equations (\ref{p4i}) 
and (\ref{phi1i}), as
\begin{equation}
\rm \frac{1}{2}F(\phi) \phi^{\prime 2}= \phi_0 e^{-2(\beta_1+\beta_2+\beta_3)}, \qquad \Rightarrow \qquad 
\sqrt{F(\phi)} d\phi=\sqrt{2\phi_0}\,e^{-(\beta_1+\beta_2+\beta_3)} d\tau, \label{solutionphi}
\end{equation}
which correspond to equation (\ref{phifield}) obtained in direct way from the original Einstein field equation. 
The corresponding classical  solutions for the field $\phi$ for this cosmological model can be seen in ref. 
\citep{socorro1}.

Using this result and the equation for the field $\phi$ given in (\ref{uno}) we can find that $\rm 2\frac{\Pi^2_\phi}{F}=16\phi_0$. 
From the hamilton
equation for the momenta $\Pi_1$ can be written for the two equations of state $\gamma=\pm 1$, introducing the generic parameter 
\begin{equation}
\rm \lambda = \left\{
\begin{tabular}{lr}
$\rm -4 \Lambda$& $\gamma=1$\\
$\rm -4 \Lambda+32\pi G \rho_1$& $\gamma=-1$
\end{tabular}
\right.
\end{equation}
as $\rm \Pi^\prime_1=\lambda e^{\beta_1+\beta_2+\beta_3}$, 
then re-introducing into the Hamiltonian equation (\ref{hami}) we find 
one differential equation for the momenta $\Pi_1$ as
\begin{equation}
\rm \frac{4}{\lambda}\Pi_1^{\prime 2}+2\Pi_1^2 -\kappa \Pi_1-k_3=0,
\end{equation}
where the corresponding constants are
\begin{equation}
\rm  \kappa=2(k_1+k_2), \quad 
k_3=\left\{ \begin{tabular}{lr}
$\rm k_1^2+k_2^2-16\phi_0$,& $\gamma=-1$\\
$\rm k_1^2+k_2^2-16\phi_0 +128 \pi G \rho_1$,& $\gamma=1$\\
\end{tabular}
\right.
\end{equation}
and whose solution is 
\begin{equation}
\rm \Pi_1=\frac{\kappa}{6}\pm \frac{\sqrt{\kappa^2+12 k_3}}{6} \, sin\left[\frac{\sqrt{3\lambda}}{2} \Delta \tau \right].
\end{equation}

On the other hand, using this result  in the sum of equation (\ref{beta1},\ref{beta2},\ref{beta3}), we obtain that
\begin{equation}
\rm \beta_1+\beta_2+\beta_3=Ln \left[\frac{\alpha}{\sqrt{12\lambda}} \, cos\left[\frac{\sqrt{3\lambda}}{2} \Delta \tau \right] \right], 
\qquad \alpha= 2\sqrt{\kappa^2+12k_3}
\end{equation}
solution previously found in ref. \citep{socorro1} using the Hamilton-Jacobi approach.
\subsection{Bianchi's Class A cosmological models }
The corresponding Lagrangian for these cosmological model are written using the Lagrangian to Bianchi I, as
\begin{eqnarray}
\rm {\cal L}_{II}&=& \rm {\cal L}_I +Ne^{\beta_1+\beta_2+\beta_3}\left[\frac{1}{2} e^{2(\beta_1-\beta_2-\beta_3)}\right],\label{ii} \\
\rm {\cal L}_{VI_{h=-1}}&=& \rm {\cal L}_I +Ne^{\beta_1+\beta_2+\beta_3}\left[2 e^{-2\beta_3}\right],\label{vi}\\
\rm {\cal L}_{VII_{h=0}}&=& \rm {\cal L}_I +Ne^{\beta_1+\beta_2+\beta_3}\left[\frac{1}{2}e^{2(\beta_1-\beta_2-\beta_3)}
+\frac{1}{2} e^{2(-\beta_1+\beta_2-\beta_3)} - e^{-2\beta_3}\right],\label{vii}\\
\rm {\cal L}_{VIII}&=& \rm {\cal L}_I +\frac{N}{2}e^{\beta_1+\beta_2+\beta_3}\left[ e^{2(\beta_1-\beta_2-\beta_3)} +
e^{2(-\beta_1+\beta_2-\beta_3)} +e^{2(-\beta_1-\beta_2+\beta_3)} \right. \nonumber\\
&& \qquad \left. -2\left(-e^{-2\beta_1} +e^{-2\beta_2}+e^{-2\beta_3} \right)\right],\label{viii}\\
\rm {\cal L}_{IX}&=& \rm {\cal L}_I +\frac{N}{2}e^{\beta_1+\beta_2+\beta_3}\left[ e^{2(\beta_1-\beta_2-\beta_3)} +
e^{2(-\beta_1+\beta_2-\beta_3)} +e^{2(-\beta_1-\beta_2+\beta_3)} \right. \nonumber \\
&& \qquad \rm \left.-2\left(e^{-2\beta_1} +e^{-2\beta_2}+e^{-2\beta_3} \right)\right],\label{ix}
\end{eqnarray} 
the momenta associated to the variables $(\beta_i,\phi)$ are the same as in equation (\ref{phi1}), 
so, the generic Hamiltonian is
\begin{equation}
\rm {\cal H}_A= \rm{\cal H}_I - \frac{1}{2}e^{-(\beta_1+\beta_2+\beta_3)} \left[U_A(\beta_1,\beta_2,\beta_3)\right],\label{h} 
\end{equation} 
where the potential term $\rm U_A(\beta_1,\beta_2,\beta_3)$ is given in table III, where A corresponds to particular Bianchi Class A models
(I,II, $\rm VI_{h=-1}$,$\rm VII_{h=0}$,VIII,IX). 
If we choose the particular gauge to the lapse function $\rm N=e^{(\beta_1+\beta_2+\beta_3)}$, the equation (\ref{h}) is much simpler, 
\begin{equation}
\rm {\cal H}_A= \rm{\cal H}_I -\frac{1}{2} \left[U_A(\beta_1,\beta_2,\beta_3)\right],\label{h-n} 
\end{equation}
where ${\cal H}_I$ is  as in equation (\ref{hami}) but without the factor $e^{-(\beta_1+\beta_2+\beta_3)}$

\begin{table}[h]
\begin{center}
\begin{tabular}{|c|c|}  \hline
{\bf  Bianchi type} & {\bf Potential} $\rm U_A(\beta_1,\beta_2,\beta_3)$ \\  \hline
I &   0 \\ \hline 
II &  $\rm e^{4\beta_1} $ \\  \hline
$\rm VI_{h=-1}$&$\rm 4e^{2(\beta_1+\beta_2)}$  \\ \hline
$\rm VII_{h=0}$ & $\rm e^{4\beta_1}+e^{4\beta_2}-2e^{2(\beta_1+\beta_2)}$ \\ \hline 
VIII & $\rm e^{4\beta_1}+e^{4\beta_2}+e^{4\beta_3}-2e^{2(\beta_1+\beta_2)}+2e^{2(\beta_1+\beta_3)}+2e^{2(\beta_2+\beta_3)}$\\ \hline 
IX &  $\rm e^{4\beta_1}+e^{4\beta_2}+e^{4\beta_3}-2e^{2(\beta_1+\beta_2)}-2e^{2(\beta_1+\beta_3)}-2e^{2(\beta_2+\beta_3)}$ \\ \hline
\end{tabular}
\vglue .5cm
Table III. \emph{Potential  $\rm U_A(\beta_1,\beta_2,\beta_3)$ for the  Bianchi Class A Models.}
\end{center}
\end{table}
The Hamilton equations, for all Bianchi Class A cosmological models are as follows
\begin{eqnarray}
\rm  \Pi_1^\prime&=& \rm -4\Lambda e^{\beta_1+\beta_2+\beta_3}+16\pi G(1-\gamma)\rho_\gamma e^{-\gamma(\beta_1+\beta_2 +\beta_3)}\nonumber\\
&&\rm \qquad +\frac{\partial}{\partial\beta_1}\left( \frac{1}{2}e^{-(\beta_1+\beta_2+\beta_3)} \left[U_A(\beta_1,\beta_2,\beta_3)\right] \right) 
\label{p1} \\
\rm  \Pi_2^\prime&=& \rm -4\Lambda e^{\beta_1+\beta_2+\beta_3}+16\pi G(1-\gamma)\rho_\gamma e^{-\gamma(\beta_1+\beta_2 +\beta_3)} \nonumber\\
&&\rm \qquad +\frac{\partial}{\partial\beta_2}\left( \frac{1}{2}e^{-(\beta_1+\beta_2+\beta_3)} \left[U_A(\beta_1,\beta_2,\beta_3)\right] \right), 
\label{p2}\\
\rm \Pi_3^\prime&=& \rm -4\Lambda e^{\beta_1+\beta_2+\beta_3}+16\pi G(1-\gamma)\rho_\gamma e^{-\gamma(\beta_1+\beta_2 +\beta_3)} \nonumber\\
&&\rm \qquad +\frac{\partial}{\partial\beta_3}\left( \frac{1}{2}e^{-(\beta_1+\beta_2+\beta_3)} \left[U_A(\beta_1,\beta_2,\beta_3)\right] \right), \label{p3}\\
\rm  \Pi_\phi^\prime &=& \rm \frac{1}{4}e^{-(\beta_1+\beta_2+\beta_3)}\, \frac{ F^\prime}{F^2  \phi^\prime}\Pi_\phi^2 , \label{p4}\\
\rm  \beta_1^\prime&=&\rm \frac{1}{4}e^{-(\beta_1+\beta_2+\beta_3)}\left[-\Pi_1 +\Pi_2+\Pi_3 \right], \label{beta1b}\\
\rm  \beta_2^\prime&=&\rm \frac{1}{4}e^{-(\beta_1+\beta_2+\beta_3)}\left[-\Pi_2 +\Pi_1+\Pi_3 \right], \label{beta2b}\\
\rm  \beta_3^\prime&=&\rm \frac{1}{4}e^{-(\beta_1+\beta_2+\beta_3)}\left[-\Pi_3 +\Pi_1+\Pi_2 \right], \label{beta3b}\\
\rm  \phi^\prime&=& \rm \frac{1}{2F}e^{-(\beta_1+\beta_2+\beta_3)}\Pi_\phi. \label{phi1a}
\end{eqnarray}

In this cosmological models, it is remarkable that the equation for the field $\phi$ (\ref{solutionphi}) is mantained for all Bianchi Class A models,
and in particular, when we use the gauge $\rm N=e^{\beta_1+\beta_2+\beta_3}$, the solutions for this field are independent of the cosmological models.

\subsection{Classical solution in the gauge $\rm N=e^{\beta_1+\beta_2+\beta_3}$, $\Lambda=0$ and $\gamma=1$}
With these initial choices, the main equations are written for this gauge as (now a dot means $\rm \frac{d}{dt}$)

\begin{eqnarray}
\rm {\cal H}_A&=&\rm \frac{1}{8}\left[-\Pi_1^2-\Pi_2^2-\Pi_3^2 +\frac{2}{F}\Pi_\phi^2+2\Pi_1 \Pi_2
+2\Pi_1 \Pi_3 +2\Pi_2 \Pi_3 -C_1\right] \nonumber\\
&& \rm \qquad  - \frac{1}{2}\left[U_A(\beta_1,\beta_2,\beta_3)\right] , \label{hami-n}
\end{eqnarray}
with $\rm C_1=128\pi G \rho_1$.

The hamilton equation, for all Bianchi Class A cosmological models are 
\begin{eqnarray}
\rm  \dot\Pi_1&=& \rm 
+\frac{\partial}{\partial\beta_1}\left( \frac{1}{2} \left[U_A(\beta_1,\beta_2,\beta_3)\right] \right) \label{p1q} \\
\rm  \dot \Pi_2&=& \rm 
+\frac{\partial}{\partial\beta_2}\left( \frac{1}{2}\left[U_A(\beta_1,\beta_2,\beta_3)\right] \right), \label{p2q}\\
\rm \dot \Pi_3&=& \rm 
+\frac{\partial}{\partial\beta_3}\left( \frac{1}{2} \left[U_A(\beta_1,\beta_2,\beta_3)\right] \right), \label{p3q}\\
\rm  \dot \Pi_\phi &=& \rm \frac{1}{4}\, \frac{ \dot F}{F^2 \dot \phi}\Pi_\phi^2 , \label{p4q}\\
\rm \dot \beta_1&=&\rm \frac{1}{4}\left[-\Pi_1 +\Pi_2+\Pi_3 \right], \label{beta1a}\\
\rm \dot \beta_2&=&\rm \frac{1}{4}\left[-\Pi_2 +\Pi_1+\Pi_3 \right], \label{beta2a}\\
\rm \dot \beta_3&=&\rm \frac{1}{4}\left[-\Pi_3 +\Pi_1+\Pi_2 \right], \label{beta3a}\\
\rm \dot \phi   &=& \rm \frac{1}{2F}\Pi_\phi. \label{phi1}
\end{eqnarray}

\subsubsection{Bianchi II}
\begin{eqnarray}
\rm  \dot\Pi_1&=& \rm 2e^{4\beta_1} \label{p1n} \\
\rm  \dot \Pi_2&=& \rm 0,  \qquad \rightarrow\qquad \Pi_2=p_2=cte, \label{p2n}\\
\rm \dot \Pi_3&=& \rm 0, \qquad \rightarrow\qquad \Pi_3=p_3=cte, \label{p3n}\\
\rm  \dot \Pi_\phi &=& \rm \frac{1}{4}\, \frac{ \dot F}{F^2  \dot \phi}\Pi_\phi^2 , \label{p4n}\\
\rm \dot \beta_1&=&\rm \frac{1}{4}\left[-\Pi_1 +p_2+p_3 \right], \label{beta1n}\\
\rm \dot \beta_2&=&\rm \frac{1}{4}\left[-p_2 +\Pi_1+p_3 \right], \label{beta2n}\\
\rm \dot \beta_3&=&\rm \frac{1}{4}\left[-p_3 +\Pi_1+p_2 \right], \label{beta3n}\\
\rm \dot \phi   &=& \rm \frac{1}{2F}\Pi_\phi. \label{phi1n}
\end{eqnarray}
introducing (\ref{p1n}) into (\ref{hami-n}) we find the differential equation for $\Pi_1$ as $\dot \Pi_1=-\frac{1}{2} \Pi^2_1 +b \Pi_1+c$ where
the constants are defined as
$\rm b=p_2+p_3$ and $\rm c=8\phi_0-\frac{1}{2}\left(p_2^2+p_3^2 + C_1 \right)$.
The solution for $\Pi_1$ is
\begin{equation}
\rm \Pi_1=b+ \sqrt{-b^2-2c} Tan \left[-\frac{1}{2} \sqrt{-b^2-2c} \Delta t\right],\label{p1-ii}
\end{equation}
and the solutions for $\beta_i$ then are
\begin{eqnarray}
\rm \Delta \beta_1&=& \rm - \frac{1}{2} \, Log\left[Cos\left(\frac{1}{2}\sqrt{-b^2-2c} \Delta t\right)\right],\\ 
\rm \Delta \beta_2&=& \rm \frac{1}{2}p_3 \Delta t + \frac{1}{2} \, Log\left[Cos\left(\frac{1}{2}\sqrt{-b^2-2c} \Delta t\right)\right]\\ 
\rm \Delta \beta_3&=& \rm \frac{1}{2}p_2 \Delta t + \frac{1}{2} \, Log\left[Cos\left(\frac{1}{2}\sqrt{-b^2-2c} \Delta t\right)\right],\\ 
\end{eqnarray}
and the solution for the $\phi$ field is similar to (\ref{solutionphi})
\begin{equation}
\rm \frac{1}{2}F(\phi) \dot \phi^2= \phi_0 , \qquad \Rightarrow \qquad 
\sqrt{F(\phi)} d\phi=\sqrt{2\phi_0}\, dt, \label{solutionphin}
\end{equation}
So, the solutions in the original variables are
\begin{eqnarray}
\rm \Omega&=& \rm \frac{1}{6}\left[ \left(p_2+p_3 \right)\Delta t +  Log\left[Cos\left(\frac{1}{2}\sqrt{-b^2-2c} \Delta t\right)\right]\right] \nonumber\\
\rm \beta_-&=& \rm \frac{\sqrt{3}}{6}\left[-\frac{1}{2}p_3 \Delta t - Log\left[Cos\left(\frac{1}{2}\sqrt{-b^2-2c} \Delta t\right)\right] \right],\nonumber\\
\rm \beta_+&=& \rm \frac{1}{12}\left[\left(p3-2p_2\right) \Delta t  -2 Log\left[Cos\left(\frac{1}{2}\sqrt{-b^2-2c} \Delta t\right)\right] \right].
\end{eqnarray}

\subsubsection{Bianchi $\rm VI_{h=-1}$}
\begin{eqnarray}
\rm  \dot\Pi_1&=& \rm 4e^{2(\beta_1+\beta_2)} \label{p1nn} \\
\rm  \dot \Pi_2&=& \rm 4e^{2(\beta_1+\beta_2)},  \qquad \rightarrow\qquad \Pi_2=\Pi_1+a_1, \label{p2nn}\\
\rm \dot \Pi_3&=& \rm 0, \qquad \rightarrow\qquad \Pi_3=p_3=cte, \label{p3nn}\\
\rm  \dot \Pi_\phi &=& \rm \frac{1}{4}\, \frac{\dot F}{F^2 \dot \phi}\Pi_\phi^2 , \label{p4nn}\\
\rm \dot \beta_1&=&\rm \frac{1}{4}\left[-\Pi_1 +\Pi_2+p_3 \right], \label{beta1nn}\\
\rm \dot \beta_2&=&\rm \frac{1}{4}\left[-\Pi_2 +\Pi_1+p_3 \right], \label{beta2nn}\\
\rm \dot \beta_3&=&\rm \frac{1}{4}\left[-p_3 +\Pi_1+\Pi_2 \right], \label{beta3nn}\\
\rm \dot \phi   &=& \rm \frac{1}{2F}\Pi_\phi. \label{phi1nn}
\end{eqnarray}
introducing (\ref{p2nn})  into (\ref{hami-n}) we find the differential equation for $\Pi_1$ as $\dot \Pi_1-p_3 \Pi_1+k_1=0$ where
$\rm k_1=\frac{1}{4}\left(p_3^2+a_1^2 -16\phi_0+ C_1-2a_1 p_3 \right)$
who solution become as
\begin{equation}
\rm \Pi_1=\frac{1}{p_3}\, \left[ e^{p_3 \Delta t} + k_1\right],\label{p1-vi}
\end{equation}
then the solutions for $\beta_i$ become
\begin{eqnarray}
\rm \Delta \beta_1&=& \rm \frac{1}{4}(a_1+p_3) \Delta t ,\\ 
\rm \Delta \beta_2&=& \rm \frac{1}{4}(p_3-a_1) \Delta t ,\\ 
\rm \Delta \beta_3&=& \rm \frac{1}{4}(a_1-p_3) \Delta t+ \frac{1}{2p_3}\, \left[ e^{p_3 \Delta t} + k_1\right],\\ 
\end{eqnarray}
and the solutions in the original variables are
\begin{eqnarray}
\Omega&=& \rm \frac{1}{12p_3}\left[2k_1+p_3\left(a_1 + p_3 \right) \Delta t +2 e^{p_3 \Delta t} \right], \nonumber\\
\beta_1&=& \rm \frac{a_1}{4\sqrt{3}}\Delta t,\nonumber\\
\beta_+&=& \rm -\frac{1}{12p_3}\left[2k_1+p_3\left(a_1 -2 p_3 \right) \Delta t +2 e^{p_3 \Delta t} \right].
\end{eqnarray}

\section{quantum scheme}
The WDW equation for these models is achived by replacing  
 $\rm \Pi_{q^\mu}=-i \partial_{q^\mu}$ in (\ref {ham}). The factor $\rm e^{-3\Omega}$ may be 
factor ordered
with $\rm \hat \Pi_\Omega$ in many ways. Hartle and Hawking \citep{HH} have 
suggested what might be called a semi-general factor ordering which in this 
case would order $\rm e^{-3\Omega} \hat \Pi^2_\Omega$ as
\begin{eqnarray}
\rm - e^{-(3- Q)\Omega}\, \partial_\Omega e^{-Q\Omega} \partial_\Omega&=&\rm
- e^{-3\Omega}\, \partial^2_\Omega + Q\, e^{-3\Omega} \partial_\Omega, \nonumber\\
 -\frac{6}{F}\phi^s \frac{\partial}{\partial \phi}\phi^{-s}\frac{\partial}{\partial \phi}&=&\rm -\frac{6}{F} \frac{\partial^2}{\partial \phi^2}+
 \frac{6s}{F}\phi^{-1}\frac{\partial }{\partial \phi}
\label {fo}
\end{eqnarray}
where  Q and s are any real constants that measure the ambiguity in the factor ordering in the variables $\Omega$ and $\phi$. 
We will assume in the following this factor ordering for the Wheeler-DeWitt equation, which becomes
\begin{equation}
\rm \Box \, \Psi -\frac{6}{F(\phi)}\frac{\partial^2 \Psi}{ \partial \phi^2}+\frac{6s}{F}\phi^{-1}\frac{\partial \Psi}{\partial \phi}
+ Q \frac{\partial \Psi}{\partial \Omega} - U(\Omega,\beta_\pm) \, \Psi -C_1\Psi =0, 
\label {WDW}
\end{equation}
where $\Box$ is the  three dimensional d'Lambertian in the $\rm \ell^\mu=(\Omega,\beta_+,\beta_-)$ 
coordinates, with signature (- + +). 

When we introduce the Ansatz 
$\rm \Psi = \chi(\phi) \psi(\Omega,\beta_\pm)$ in (\ref {WDW}), we obtain the general set of differential
equations (under the assumed factor ordering) for the Bianchi type IX cosmological model
\begin{eqnarray}
\rm \Box \, \psi + Q \frac{\partial \psi}{\partial \Omega} - \left[U(\Omega,\beta_\pm)+C_1-\mu^2\right] \, \psi &=&\rm 0,\label{wdwmod}\\
\rm \frac{6}{F(\phi)}\frac{\partial^2 \chi}{ \partial \phi^2}-\frac{6s}{F}\phi^{-1}\frac{\partial \chi}{\partial \phi} +\mu^2 \chi&=&0 
\label {phi-1}
\end{eqnarray}
 
When we calculate the solution to equation (\ref{phi-1}), we find interesting properties on this, as
\begin{enumerate}
\item{} This equation is a master equation for the field $\phi$ for any cosmological model, implying that this field $\phi$ is an universal field
as cosmic ground, having the best presence in the stiff matter era as an ingredient in the formation the structure galaxies and when we consider 
two types of functions, $\rm F(\phi)=\omega \phi^m$ and $\rm F(\phi)=\omega e^{m\phi}$, we have the following exact solutions \citep{andrei} 
\begin{enumerate}
\item{$F(\phi)=\omega\phi^m$}\\
the differential equation to solver is
\begin{equation}
\frac{d^2\chi}{d \phi^2 }-s\phi^{-1}\frac{d\chi}{d \phi}+\alpha\phi^m\chi=0 \label{cphi0}
\end{equation}
with $\alpha= \frac{\omega \mu^2}{6}$. The solutions depend on the value to $m$ and $s$,

\begin{enumerate}
\item{} General solution for any $m\not=-2$ and $s$, are written in terms of ordinary and modify Bessel function,
\begin{equation}
 \chi=c_1 \phi^{\frac{1+s}{2}}\, Z_\nu \left(\frac{2\sqrt{\alpha}}{m+2} \phi^{\frac{m+2}{2}} \right),
\end{equation}
with  $c_1$ an integration constant,  $ Z_\nu$ is a generic Bessel function, $\nu=\frac{1+s}{m+2}$ is the order. 
When $\alpha>0$ imply  $\omega>0$, $ Z_\nu$ become the ordinary Bessel function, $(J_\nu, Y_\nu)$. If $\alpha<0, \to w<0$, 
$ Z_\nu \to (I_\nu,K_\nu)$.
\item{$m=-2$ and any $s$, }
\begin{equation}
\chi=\phi^{\frac{1+s}{2}} \left\{
\begin{tabular}{lr}
$  c_1\, \phi^\mu + c_2 \phi^{-\mu} $ & $\qquad$ si $\quad \mu>0$ \cr
$  c_1 \, + c_2 Ln \phi $ & \qquad si \quad $\mu=0$ \cr
$ c_1\, sin\left(\mu Ln \phi \right) + c_2\, cos\left(\mu Ln \phi \right)$& \qquad if \quad $\mu<0$ \cr
\end{tabular}
\right. 
\end{equation}
where $\mu=\frac{1}{2} \sqrt{|(1+s)^2-4\alpha|}$

\item{ $m=-6$ and $s=1$}
\begin{equation}
\chi(\phi)= \phi^2\left\{
\begin{tabular}{lr}
$c_1 \, sinh\left(\frac{\sqrt{|\alpha|}}{2\phi^2}\right) + c_2 \,cosh\left(\frac{\sqrt{|\alpha|}}{2\phi^2}\right)$&\quad $\alpha<0\to\omega<0$\\
$c_1 \, sin\left(\frac{\sqrt{|\alpha|}}{2\phi^2}\right) + c_2 \,cos\left(\frac{\sqrt{|\alpha|}}{2\phi^2}\right)$&\quad $\alpha>0\to\omega>0$\\
\end{tabular} \right. \label{cphi2}
\end{equation}

\end{enumerate}

\item{$F(\phi)=\omega e^{m\phi}$}, for this case we consider the caso $s=0$,
\begin{equation}
\frac{d^2\chi}{d \phi^2 }+\alpha e^{m\phi}\chi=0
\end{equation}
\begin{enumerate}
\item{$m\neq0$}
\begin{equation}
\chi=C Z_0\left(\frac{2\sqrt{\alpha}}{m}e^{\frac{m\phi}{2}}\right)
\end{equation}
with $C$ is a integration constant  and $Z_0$ is the generic Bessel function to zero order. So, if
 $\alpha>0$ then  $\omega>0$, $ Z_0$ is the ordinary Bessel function $(J_0, Y_0)$. When $\alpha<0, \to \omega<0$, 
$ Z_0 \to (I_0,K_0)$.
\item{for $m=0$},
\begin{equation}
\chi=\left\{
\begin{tabular}{cl}
$c_1\sinh\left(\sqrt{|\alpha|}\phi\right)+c_2\cosh\left(\sqrt{|\alpha|}\phi\right)$ &\quad if $\alpha<0 \to  \omega<0$\\
$c_1 \sin\left(\sqrt{|\alpha|}\phi\right)+c_2 \cos\left(\sqrt{|\alpha|}\phi\right)$ &\quad if $\alpha>0 \to  \omega>0$\\
\end{tabular}
\right.
\end{equation}
\end{enumerate}
\end{enumerate}
\item{} If we have the solution for the parameter s=0 for arbitrary
function $\rm F(\phi)$, say $\chi_0$, then we have also the solution for s=-2, as $\chi(s=-2)=\frac{\chi_0}{\phi}$.
\end{enumerate}

To obtain the solution of the other factor of $\Psi$ we use  the particular value for the constants
$\rm C_1=\mu^2$, and make the following Ansatz for the wave function
\begin{equation}
\rm \psi(\ell^\mu) = W(\ell^\mu) e^{- S(\ell^\mu)}, \label{ans}
\end{equation}
 where $S(\ell^\mu)$ is known as the superpotential function, and W is the amplitude of probability to that employed in Bohmian
  formalism \citep{bohm}, those found in the literature, years ago \citep{os}. So 
(\ref {wdwmod}) is transformed into    
\begin{equation}
 {\Box \, W} - W {\Box \, S} - 2 {\nabla W}\cdot {\nabla S} + 
Q \frac{\partial W}{\partial \Omega} - Q W \frac{\partial S}{\partial \Omega}+
 W \left[ \left(\nabla S\right)^2 - U\right] = 0, 
\label {mod}
\end{equation}
where  $\rm \Box = G^{\mu \nu}\frac{\partial^2}{\partial \ell^\mu \partial \ell^\nu}$,
$\rm {\nabla \, W}\cdot {\nabla \, \Phi}=G^{\mu \nu}
\frac{\partial W}{\partial \ell^\mu}\frac{\partial \Phi}{\partial \ell^\nu}$,
$\rm (\nabla)^2= G^{\mu \nu}\frac{\partial }{\partial \ell^\mu}\frac{\partial }{\partial \ell^\nu}=
-(\frac{\partial}{\partial \Omega})^2 +(\frac{\partial}{\partial \beta_+})^2 +
(\frac{\partial}{\partial \beta_-})^2$, with 
$\rm G^{\mu \nu}= diag(-1,1,1)$,  U is the
potential term of the cosmological model under consideration. 

 Eq  (\ref{mod}) 
can be written as the following set of partial differential equations
\begin{subequations}
\label{WDWa}
\begin{eqnarray}
(\nabla S)^2 - U &=& 0, \label{hj} \\
  W \left( \Box S + Q \frac{\partial S}{\partial \Omega}
  \right) + 2 \nabla \, W \cdot \nabla \, S &=& 0 \, , \label{wdwho} \\
  \Box \, W + Q \frac{\partial W}{\partial \Omega} & = & 0 \label{cons}  \, . 
\end{eqnarray}
\end{subequations}

Following reference \citep{wssa}, first we shall choose to solve 
Eqs. (\ref{hj}) and (\ref{wdwho}), whose solutions at the end will have to fulfill Eq. (\ref{cons}),  
which play the role of a constraint equation.

\subsection{Transformation of the Wheeler-DeWitt equation}

We were able to solve (\ref {hj}), by doing
the change of coordinates (\ref{transformation}) 
 and rewrite (\ref {hj})
in these new coordinates. With this change, the function S is obtained
as follow, with the ansatz (\ref{ans}),

In this section, we obtain the solutions to the equations that appear in
the decomposition of the WDW equation,  
(\ref {hj}), (\ref {wdwho}) and (\ref {cons}), using the Bianchi type IX Cosmological model.
So, the equation $\rm [\nabla]^2= -(\frac{\partial}{\partial \Omega})^2 +(\frac{\partial}{\partial \beta_+})^2
+(\frac{\partial}{\partial \beta_-})^2$ can be written in the following way (see appendix section 9)
\begin{eqnarray}
\left[\nabla \right]^2 &=& 3 \left[ \left(\frac{\partial}{\partial \beta_1}\right)^2+
\left(\frac{\partial}{\partial \beta_2} \right)^2+
\left(\frac{\partial}{\partial \beta_3}\right)^2 \right ] - 6\left[
 \frac{\partial}{\partial \beta_1} \frac{\partial}{\partial \beta_2} +
 \frac{\partial}{\partial \beta_1} \frac{\partial}{\partial \beta_3} +
 \frac{\partial}{\partial \beta_2} \frac{\partial}{\partial \beta_3}
\right ]\nonumber\\
 &=& 3  \left ( \frac{\partial}{\partial \beta_1}+
                   \frac{\partial}{\partial \beta_2}+
                   \frac{\partial}{\partial \beta_3} \right )^2 -12 \left [
\frac{\partial}{\partial \beta_1} \frac{\partial}{\partial \beta_2} +
\frac{\partial}{\partial \beta_1} \frac{\partial}{\partial \beta_3} +
\frac{\partial}{\partial \beta_2} \frac{\partial}{\partial \beta_3}
\right ].
\label {nab}
\end{eqnarray}
The potencial term of the Bianchi type IX is transformed in the new variables into
\begin{equation}
\rm U = 12 \left [ \left ( e^{2\beta_1}+e^{2\beta_2}+e^{2\beta_3} \right )^2 
-2 e^{2(\beta_1+ \beta_2)} -2 e^{2(\beta_1+ \beta_3)}-2 e^{2(\beta_2+ \beta_3)} \right ]
\label {pot}
\end{equation}
Then (\ref {hj}) for this models is rewritten  in the new 
variables as
\begin{eqnarray}
 & & 3  \left ( \frac{\partial S}{\partial \beta_1}+
               \frac{\partial S}{\partial \beta_2}+
               \frac{\partial S}{\partial \beta_3} \right)^2 -12 \left[
\frac{\partial S}{\partial \beta_1} \frac{\partial S}{\partial \beta_2} +
\frac{\partial S}{\partial \beta_1} \frac{\partial S}{\partial \beta_3} +
\frac{\partial S}{\partial \beta_2} \frac{\partial S}{\partial \beta_3}\right] \nonumber\\
& &\mbox{} -12 \left [ \left ( e^{2\beta_1}+e^{2\beta_2}+e^{2\beta_3} \right )^2 
-4 e^{2(\beta_1+ \beta_2)} -4 e^{2(\beta_1+ \beta_3)}-4 e^{2(\beta_2+ \beta_3)} \right ] =0.
\label {hanv}
\end{eqnarray}
Now, we can use the separation of variables method to get
solutions to the last equation for the $\rm S$ function, obtaining for the
Bianchi type IX model 
\begin{equation}
\rm S_{IX}= \pm \left ( e^{2\beta_1} + e^{2\beta_2} + e^{2\beta_3} \right ). 
\label {phi9}
\end{equation}
In table IV we present the corresponding superpotential function S and amplitude W  for all Bianchi Class A models.

With this result, and using for the solution to (\ref {wdwho}) in the new coordinates $\beta_i$, we have  for W 
function as
\begin{equation}
\rm  W_{_{IX}}= W_0 \, e^{ \left[(1+ \frac{Q}{6})\left(\beta_1+\beta_2+\beta_3 \right)\right]}.
\label {w9}
\end{equation}
and  re-introducing this result into  Eq. (\ref{cons}) we find that $\rm Q=\pm 6$. Therefore   we have two wave functions 
\begin{eqnarray}
\rm  \psi_{_{IX}}(\beta_i)&=&\rm W_{_{IX}}(\beta_i) \,  Exp\left[\pm \left(e^{2\beta_1}+e^{2\beta_2}+e^{2\beta_3} \right) \right]\nonumber\\
&=&\rm  Exp\left[\pm \left(e^{2\beta_1}+e^{2\beta_2}+e^{2\beta_3} \right) \right]
\left\{
\begin{tabular}{ll}
$\rm W_0,$ & \qquad Q=-6,\\
$\rm W_0 Exp\left[2\left(\beta_1+\beta_2+\beta_3 \right)\right]$,& \qquad Q=6\\
\end{tabular}
\right.
\label {psi}   \,
\end{eqnarray}
similar solutions were given by Moncrief and Ryan \citep{moncrief} in standard quantum cosmology in general relativity. In table IV 
we present the superpotential function S, the amplitude of probability W and the relations between the parameters for
the corresponding Bianchi Class A models.

\begin{table}[h]
\begin{center}
\begin{tabular}{|l|l|l|l|}   \hline
{\bf  Bianchi} & {Superpotential ${\bf S}$} & {Amplitude of probability $\rm W$}& Constraint\\ 
{\bf  type} & & & \\ \hline 
I  &  constant & $\rm e^{(\frac{r}{3}+\frac{b}{6}+\frac{\sqrt{3}c}{6})\beta_1 +(\frac{r}{3}+\frac{b}{6}-\frac{\sqrt{3}c}{6})\beta_2 +
(\frac{r}{3}-\frac{b}{3})\beta_3}$ & $r^2-Qr-a^2=0,$\\ 
& & &$\rm a^2=b^2+c^2$\\  \hline
II &  $\rm e^{2\beta_1}$ & $\rm e^{(a-1-\frac{Q}{6})\beta_1+a\beta_2+(a-b)\beta_3}$& $\rm 144b^2-144ab+36$\\
 & & &$\rm -Q^2+24aQ=0$\\ \hline
$\rm VI_{h=-1}$ & $\rm 2 (\beta_1-\beta_2)\, e^{(\beta_1+\beta_2)}$ &$\rm e^{a(\beta_1+\beta_2)}$ &$Q=0$\\ \hline
$\rm VII_{h=0}$ &  $\rm e^{2\beta_1}+ e^{2\beta_2}$ &
$\rm e^{\left(1+\frac{Q}{6}\right)\left(\beta_1+\beta_2+\beta_3 \right)+a(\beta_1+\beta_2) }$ &
$\rm Q^2-48a-36=0$\\ \hline
VIII  & $\rm    e^{2\beta_1}+ e^{2\beta_2} - e^{2\beta_3}$ &
$\rm W_0 \, e^{ \left[(1+ \frac{Q}{6})\left(\beta_1+\beta_2+\beta_3 \right)\right]}$ & $\rm Q=\pm 6$\\ \hline
IX  &$\rm  e^{2\beta_1}+ e^{2\beta_2} + e^{2\beta_3}$ 
& $\rm W_0 \, e^{ \left[(1+ \frac{Q}{6})\left(\beta_1+\beta_2+\beta_3 \right)\right]}$ & $\rm Q=\pm 6$\\
\hline
\end{tabular}
\vglue .5cm
Table IV.  \emph{superpotential $\rm S$, the amplitude of probability W and the  relations between the
parameters   for the corresponding  Bianchi Class A models.}
\end{center}
\end{table}
If one looks at the expressions for the functions S given in table IV, one notes that there is  a
general form to write them using the 3x3 matrix $\rm m^{ij}$  that appear in the classification scheme of
Ellis and MacCallum \citep{ellis} and Ryan and Shepley \citep{ryan-she}, 
the structure constants are written in the form
\begin{equation}
\rm C^i_{jk}=\epsilon_{jks} \, m^{si} + \delta^i_{[k} a_{j].}
\label {estr}
\end{equation}
where $\rm a_i=0$ for the Class A models.

If we define $\rm g_i(\beta_i)= (e^{\beta_1}, e^{\beta_2}, e^{\beta_3}),$ with 
$\rm \beta_i$ given in (\ref {transformation}), the  solution to (\ref {hj})
can be written as
\begin{equation}
\rm S(\beta_i)= \pm  [g_i \, M^{ij} \, (g_j)^T].
\label {phig}
\end{equation}  
where $\rm M^{ij}=m^{ij}$ for the Bianchi Class A, excepting the 
Bianchi type $\rm VI_{h=-1}$ for which we redefine the matrix to be 
consistent with (\ref {phig})
$$\rm  M^{ij} = \left(\beta_1-\beta_2 \right) \left(
\begin{tabular}{lcr} 
0 & 1 & 0 \\
1 & 0 & 0 \\ 
0 & 0 & 0
\end{tabular}
\right)
$$

Then, for the Bianchi Class A models, the wave function $\Psi$ can be written
in the general form
\begin{equation}
\rm \Psi= \chi(\phi) \, W(\beta_i) \, exp\,[\pm [g_i \, M^{ij} \, (g_j)^T]].
\label {psig}
\end{equation}

\section{Final remarks}
Using the analytical procedure of hamilton equation of classical mechanics, in appropriate
coordinates, we found a master equation for all Bianchi Class A cosmological models, we present
partial result in the classical regime for three models of them, but the general equation are shown for all them.
In particular, the Bianchi type I is complete solved without using
a particular gauge.
 The Bianchi type II and $\rm VI_{h=-1}$ are solved introducing a particular gauge.
An important results yields when we use the gauge $\rm N=e^{\beta_1+\beta_2+\beta_3}$, we find that the solutions for the $\phi$ 
field are independent of the cosmological models, and we find that the energy density associated has a scaling behaviors under the analysis of
 standard field theory to scalar fields \citep{andrew,ferreira}, is say, scales
exactly as a power of the scale factor like,  $\rho_\phi\propto a^{-m}$. More of this can be seen to references cited before.
 On the other hand, in the quantum regime, wave functions of the form $\rm \Psi= W\, e^{\pm S}$ are the only
known exact solutions for the Bianchi type IX model in standard quantum
cosmology. In the SB formalism, these solutions are modified only for the function 
$\rm \chi$, $\rm \Psi= \chi(\phi)\,W(\ell^\mu)\, e^{\pm S(\ell^\mu)}$ when we include the particular ansatz $\rm C_1=\mu^2$. 
This kind of solutions already have been found in supersymmetric 
quantum cosmology \citep{asano} and also
for the WDW equation defined in the bosonic sector of the heterotic strings \citep{lidsey}.
Recently, in the books \citep{paulo-vargas} 
appears all solutions in the supersymmetric scheme similar at our formalism.
We have shown that they are also exact solutions to the rest of the
 Bianchi Class A models in SB quantum cosmology, under the assumed
semi-general factor ordering (\ref {fo}). Different 
procedures seem to produce this particular quantum state, where S is
a solution to the corresponding classical Hamilton-Jacobi 
equation (\ref {hj}). 

\section{Appendix: Energy momentum tensor}
From Eq. (\ref{ener-mom}) we see that the effective energy momentum tensor of the scalar field is 
\begin{equation}
T_{\alpha \; \beta}=F(\phi) \left( \phi_{,\alpha}\phi_{,\beta}
 - \frac{1}{2} g_{\alpha \beta} \phi_{,\gamma} \phi^{,\gamma} \right)
\end{equation}
this energy momentum tensor is conserved, as follows from the equation of motion for the scalar field
\begin{eqnarray}
\rm \nabla^\beta T_{\alpha \; \beta}&=&\rm   \nabla^\beta \left[F(\phi) \left( \phi_{,\alpha}\phi_{,\beta}
 - \frac{1}{2} g_{\alpha \beta} \phi_{,\gamma} \phi^{,\gamma} \right)\right]=
 F'(\phi) \phi^{,\beta} \left( \phi_{,\alpha}\phi_{,\beta}
 - \frac{1}{2} g_{\alpha \beta} \phi_{,\gamma} \phi^{,\gamma} \right) \nonumber \\
&& \qquad \rm +
F(\phi) \left( \phi_{,\alpha}^{\;\; ;\beta}\phi_{,\beta}+ \phi_{,\alpha}\phi_{,\beta}^{\;\; ;\beta}
 - \frac{1}{2} g_{\alpha \beta} \phi_{,\gamma}^{\;\; ;\beta} \phi^{,\gamma}  - \frac{1}{2} g_{\alpha \beta} \phi_{,\gamma} \phi^{,\gamma ;\beta}\right) \nonumber \\
&=&  \rm F'(\phi) \left(  \frac{1}{2}  \phi_{,\gamma} \phi^{,\gamma} \phi_{,\alpha} \right)+
F(\phi) \left( \phi_{,\alpha}^{\;\; ;\beta}\phi_{,\beta}+ \phi_{,\alpha}\phi_{,\beta}^{\;\; ;\beta}
 -  g_{\alpha \beta} \phi_{,\gamma}^{\;\; ;\beta} \phi^{,\gamma} \right) \nonumber \\
&=& \rm \frac{1}{2} \phi_{,\alpha} \left(  F'(\phi)  \phi_{,\gamma} \phi^{,\gamma} +2 F(\phi)   \phi_{,\alpha}\phi_{,\beta}^{\;\; ;\beta} \right)=0
\end{eqnarray}
Now we proceed to show that the energy momentum tensor has the structure of an imperfect  stiff fluid,
\begin{equation}
\rm T_{\alpha \; \beta}= (\rho+p) U_{\alpha } U_{ \beta} +p  g_{\alpha \; \beta}= (2 \rho) [ U_{\alpha } U_{ \beta} +\frac{1}{2} g_{\alpha \; \beta}]
\end{equation}
here $\rho$ is the energy density, $p$ the pressure, and   $ U_{\alpha }$ the velocity
If we choose for the velocity the normalized derivative of the scalar field, assuming that it is a timelike vector, 
as is often the case in cosmology, where the scalar field is only time dependent

\begin{equation}
 \rm U_{\alpha }=S^{-1/2} \phi_{,\alpha } ,   \qquad S= - \phi_{,\sigma } \phi^{,\sigma } ,
\end{equation}
It is evident that the energy momentum tensor of the SB theory is equivalent to a stiff fluid with the energy density given by

\begin{equation}
\rm \rho=  \frac{S\, F(\phi) }{2}  = -   \frac{  \phi_{,\sigma } \phi^{,\sigma }    \, F(\phi) }{2}.
\end{equation}
Therefore the most important contribution of the scalar field occurs during a stiff matter phase that is previous to the  dust phase.

\section{Appendix: Operators in the $\beta_i$ variables}
The operators who appear in eqn (\ref{WDW}) are calculated in the original variables $\rm (\Omega, \beta_+,\beta_-)$; however the structure 
of the cosmological potential term gives us an idea to implement new variables, considering the Bianchi type IX cosmological model, these one given
by eqn (\ref{transformation}). The main calculations are based in the following
\begin{eqnarray}
\rm \frac{\partial}{\partial \Omega}&=& \rm \frac{\partial}{\partial \beta_1}+\frac{\partial}{\partial \beta_2}+\frac{\partial}{\partial \beta_3},\nonumber\\
\rm \frac{\partial^2}{\partial \Omega^2}&=& \rm \frac{\partial^2}{\partial \beta_1^2}+\frac{\partial^2}{\partial \beta_2^2}
+\frac{\partial^2}{\partial \beta_3^2}+ 2\left[\frac{\partial^2}{\partial \beta_1 \partial \beta_2}
+\frac{\partial^2}{\partial \beta_1 \partial \beta_3}+\frac{\partial^2}{\partial \beta_2 \partial \beta_3} \right], \nonumber\\
\rm \frac{\partial}{\partial \beta_+}&=& \rm \frac{\partial}{\partial \beta_1}+\frac{\partial}{\partial \beta_2}-2\frac{\partial}{\partial \beta_3},\nonumber\\
\rm \frac{\partial^2}{\partial \beta_+^2}&=& \rm \frac{\partial^2}{\partial \beta_1^2}+\frac{\partial^2}{\partial \beta_2^2}
+4\frac{\partial^2}{\partial \beta_3^2}+ 2\left[\frac{\partial^2}{\partial \beta_1 \partial \beta_2}
-2\frac{\partial^2}{\partial \beta_1 \partial \beta_3}-2\frac{\partial^2}{\partial \beta_2 \partial \beta_3} \right], \nonumber\\
\rm \frac{\partial}{\partial \beta_-}&=& \rm \sqrt{3}\left(\frac{\partial}{\partial \beta_1}-\frac{\partial}{\partial \beta_2}\right),\nonumber\\
\rm \frac{\partial^2}{\partial \beta_-^2}&=&3 \left(\rm \frac{\partial^2}{\partial \beta_1^2}+\frac{\partial^2}{\partial \beta_2^2}-
2\frac{\partial^2}{\partial \beta_1 \partial \beta_2}\right).
\end{eqnarray}
So, the operator $(\nabla)^2$, $\Box$, $\nabla S \nabla W$ are written as
\begin{eqnarray}
\rm (\nabla)^2&=& \rm G^{\mu\nu} \frac{\partial }{\partial \ell^\mu} \frac{\partial }{\partial \ell^\nu}, \qquad G^{\mu \nu}= diag(-1,1,1),
\qquad \ell^\mu=(\Omega, \beta_+,\beta_1),\nonumber\\
&=&\rm 3\left\{ \left( \frac{\partial}{\partial \beta_1} \right)^2 +\left(\frac{\partial}{\partial \beta_2} \right)^2
+\left(\frac{\partial}{\partial \beta_3} \right)^2-2\left[\frac{\partial}{\partial \beta_1} \frac{\partial}{\partial \beta_2}
+\frac{\partial}{\partial \beta_1} \frac{\partial}{\partial \beta_3} 
+ \frac{\partial}{\partial \beta_2} \frac{\partial}{\partial \beta_3}  \right] \right\}\nonumber\\
&=&\rm 3\left\{ \left[\frac{\partial}{\partial \beta_1}  +\frac{\partial}{\partial \beta_2} 
+\frac{\partial}{\partial \beta_3} \right]^2-4\left[\frac{\partial}{\partial \beta_1} \frac{\partial}{\partial \beta_2}
+\frac{\partial}{\partial \beta_1} \frac{\partial}{\partial \beta_3} 
+ \frac{\partial}{\partial \beta_2} \frac{\partial}{\partial \beta_3}  \right] \right\}\nonumber\\
\rm \Box&=& \rm  G^{\mu \nu}\frac{\partial^2}{\partial \ell^\mu  \partial \ell^\nu}
= 3\left(\frac{\partial^2}{\partial \beta_1^2}+\frac{\partial^2}{\partial \beta_2^2}
+\frac{\partial^2}{\partial \beta_3^2} \right) - 6 \left( \frac{\partial^2}{\partial \beta_1 \partial \beta_2}
+\frac{\partial^2}{\partial \beta_1 \partial \beta_3}+\frac{\partial^2}{\partial \beta_2 \partial \beta_3}\right)\nonumber\\
\rm \nabla S \cdot \nabla W&=& \rm G^{\mu\nu} \frac{\partial S}{\partial \ell^\mu} \frac{\partial W}{\partial \ell^\nu}, \nonumber\\
&=& \rm 3\left(\frac{\partial S}{\partial \beta_1}\frac{\partial W}{\partial \beta_1}
+\frac{\partial S}{\partial \beta_2}\frac{\partial W}{\partial \beta_2}
+\frac{\partial S}{\partial \beta_3}\frac{\partial W}{\partial \beta_3}  \right) \nonumber\\
&& \rm -
3\left(\frac{\partial S}{\partial \beta_1}\frac{\partial W}{\partial \beta_2} +\frac{\partial S}{\partial \beta_1}\frac{\partial W}{\partial \beta_3}
+\frac{\partial S}{\partial \beta_2}\frac{\partial W}{\partial \beta_3} +\frac{\partial S}{\partial \beta_2}\frac{\partial W}{\partial \beta_1}
+\frac{\partial S}{\partial \beta_3}\frac{\partial W}{\partial \beta_1} +\frac{\partial S}{\partial \beta_3}\frac{\partial W}{\partial \beta_2} \right)
\end{eqnarray}

\section{acknowledgments}
This work was partially supported by CONACYT grant 56946. 
DAIP (2010-2011) and PROMEP grants UGTO-CA-3. This work is part of the collaboration within the Instituto Avanzado de
Cosmolog\'{\i}a. Many calculations were done by Symbolic Program REDUCE 3.8.

\end{document}